\documentclass[9pt,conference]{IEEEtran}

\def\br{{\boldsymbol{r}}}
\def\ba{{\boldsymbol{a}}}
\def\btau{{\boldsymbol{\tau}}}
\def\bd{{\boldsymbol{d}}}
\def\bchi{{\boldsymbol{\chi}}}
\def\bg{{\boldsymbol{g}}}

\def\bI{{\boldsymbol{I}}}
\def\bS{{\boldsymbol{S}}}
\def\bA{{\boldsymbol{A}}}
\def\bG{{\boldsymbol{G}}}

\def\bD{{\boldsymbol{D}}}
\def\bTheta{{\boldsymbol{\theta}}}
\def\bmu{{\boldsymbol{\mu}}}

\def\btau{\boldsymbol \tau}

\usepackage{amsbsy,amsfonts,amsfonts,amssymb,color,amsmath,psfrag}
\usepackage{graphics}
\usepackage{graphicx}
\begin{document}
\newcommand{\exd}[2]{\tilde{d}_{#1}^{[i]}(#2)}
\newcommand{\corrd}[4]{\left(\widetilde{d_{#1}(#2) d_{#3}(#4)}\right)^{[i]}}
\def\thefootnote{}

\title{\Huge  A Monte-Carlo Implementation of the SAGE Algorithm for Joint Soft Multiuser and Channel Parameter Estimation}
\author{\Large \authorblockN{E. Panayirci \authorrefmark{1}, A. Kocian
    \authorrefmark{2}, H. V. Poor \authorrefmark{1}, and M. Ruggieri \authorrefmark{2}}
\authorblockA{ \authorrefmark{1} Department of Electrical Engineering, Princeton
University, Princeton, NJ 08544, USA}
\authorblockA{ \authorrefmark{2} University of Rome "Tor Vergata",
Center for TeleInFrastructure (CTiF)-Italy, Rome, Italy}
}

\maketitle

\footnotetext{\hspace*{-3ex} E. Panayirci is on
sabbatical leave from Kadir Has University, Istanbul
Turkey.\\
This research has been conducted in part within the NEWCOM++
Network of Excellence in Wireless Communications  project funded
through the EC 7th Framework Program, and in part under the support of the U.S. National Science Foundation under Grant CNS-06-25637.}

\thispagestyle{empty}

\begin{abstract}

An efficient, joint transmission delay and
channel parameter estimation algorithm is proposed for  uplink
asynchronous direct-sequence code-division multiple access (DS-CDMA)
systems based on the space-alternating generalized expectation
maximization (SAGE) framework. The marginal likelihood of the unknown
parameters, averaged over the data sequence, as well as the
expectation and maximization steps of the SAGE algorithm  are
derived analytically. To implement the proposed
algorithm, a Markov Chain Monte Carlo (MCMC) technique,
called Gibbs sampling, is employed to compute the {\em a
posteriori} probabilities of data symbols in a computationally
efficient way. 
Computer simulations show that the proposed algorithm
has excellent estimation performance. This so-called MCMC-SAGE
receiver is guaranteed to converge in likelihood.\\

{\bf Index Terms-} Asynchronous DS-CDMA, space-alternating
generalized expectation maximization(SAGE), Markov Chain Monte Carlo
(MCMC), Gibbs sampling.

\end{abstract}

\section{Introduction}

 The performance of direct-sequence code-division multiple-access (DS-CDMA)
transmission over mobile fading channels depends strongly on the
reliability of channel parameter and quality of synchronization for each user: state-of-the-art detection algorithms that
exploit multiple-access-interference and inter-symbol-interference
require very powerful estimation algorithms.

Substantial amount of relevant references appeared in the
literature on  delay estimation. Namely, a new prospective is
presented in \cite{stuller97} for the maximum likelihood (ML) time-delay estimation. Code timing estimation in a near-far environment for DS-CDMA systems was introduced in \cite{smith}.
 Joint symbol detection, time-delay and channel parameter estimation
problems for asynchronous DS-CDMA systems have been investigated in several previous works (e.g., \cite{pelin,miller}). Most of these works either work on one signal at a time and treat the other signals as interference, or employ a training sequence to obtain a coarse
estimate of the channel parameters which is consequently used to
detect data. It is clear that these approaches have disadvantages of
having higher overhead and additional noise enhancement.

Some other proposed approaches for joint blind multiuser detection and
channel estimation for DS-CDMA systems are  subspace-based  and
linear prediction-based methods. Subspace-based method usually
require singular value decomposition or eigenvalue decomposition
which is computationally costly and does not tolerate mismatched
channel parameters. Another drawback of this approach is that
accurate rank determination may be difficult in a noisy
environment \cite{bensley,strom}. Moreover, it is not clear how these methods can be extended to include the estimation of the transmission delays jointly with the channel parameters.

The expectation maximization (EM) and space alternating EM (SAGE) algorithms are
ideally suited to these kind of problems as they are guaranteed to
converge in likelihood. Earlier  work related with delay estimation based on the EM algorithm  has appeared \cite{Georg-90,Georg-91}.
Efficient iterative receiver structures are presented in
\cite{kocian2003,kocian2007}, performing joint multiuser detection and
channel estimation  for synchronous as well as asynchronous coded
DS-CDMA systems operating over quasi-static flat Rayleigh fading
channels, under the assumption that the transmissions delays are
known. The Bayesian EM/SAGE algorithm can be used for joint
\emph{soft}-data estimation and channel estimation but the
computational complexity of the resulting receiver architecture is non-polynomial in the number of
users \cite{Gallo-04}. To overcome this draw-back, Hu \emph{et al.}
applied the Variational Bayesian EM/SAGE algorithms to joint
estimation of the distributions for channel coefficients, noise
variance, and information symbols for synchronous DS-CDMA in
\cite{Hu-08}. Our work may be considered to be a twofold extension of the
work by Gallo \emph{et al.} in \cite{Gallo-04}:
First, the proposed receiver performs joint channel coefficient and
transmission delay estimation within the SAGE framework. Secondly, the
implication of the Monte-Carlo method in the SAGE framework makes it
possible to compute \emph{soft}-data estimates for all users at polynomial
computational complexity, as well. Here, an
efficient Markov chain Monte Carlo (MCMC) technique \cite{gelfand} called
{\em Gibbs sampling} is used to compute the {\em a posteriori}
probabilities (APP) of data symbols \cite{doucet}. The APP's can be computed
exactly with the MCMC algorithm, which is significantly less complex than
a standard hidden Markov model approach. 
The resulting receiver architecture
works in principal fully blind and is guaranteed to converge. For
uncoded transmission, a few
pilot bits must be inserted, though, to resolve the phase ambiguity problem.

The theoretical framework for the joint transmission delays and
channel estimation as well as the data detection algorithms can easily
be extended to coded transmission.

\section {System Description}
\label{sec:system}

We consider an asynchronous single-rate DS-CDMA system with $K$
active users using binary phase shift keying (BPSK) modulation
sharing the same propagation channel. The signal transmitted by each
user experiences flat Rayleigh fading, which is assumed to be
constant over the observation frame of $L$ data symbols. Each user employs
a random signature  waveform for transmitting symbols of duration
$T_{b}$, such that each symbol consists of $N_{c}$ chips with duration
$T_{c}=T_{b}/N_{c}$ where $N_{c}$ is an integer. The received signal
is the noisy sum of all user's contribution, delayed by the propagation delays $\tau_{k}\in [0,T_{b}/2)$, where the subscript $k$
denotes the label of  the $k$th user. After down-converting the
received signal to baseband and passing it through an
integrate-and-dump filter with integration time $T_{s}=T_{c}/Q$, $Q
\in \mathbb{Q}^+$, $QN_c(L+1)$ samples over an observation frame of $L$ symbols  are
stacked into a signal column vector $\br \in \mathbb{C}^{Q N_{c}
(L+1)-1}$. Note that sampling is chip-synchronous without
knowledge of the individual transmission delays. It can
therefore be expressed as
\begin{equation}
\boldsymbol{r} = 
\boldsymbol{S}(\boldsymbol{\tau})\bf{A}\bf{d}+\boldsymbol{w}.
\label{sys:received}
\end{equation}

In this expression the matrix
$\boldsymbol{S} (\boldsymbol{\tau})
\in \mathbb{C}^{Q N_{c} (L+1)-1 \times LK}$ contains the signature
sequences of all the users
\[
\boldsymbol{S}( \btau)= \left[ {\begin{array}{*{20}c}
   \boldsymbol{S}_{1} (\tau_1), &
   \boldsymbol{S}_{2}(\tau_2),  & \cdots &  ,\boldsymbol{S}_{K} (\tau_K) \\
\end{array}} \right]
\]
where $\boldsymbol{S}_k (\tau_k) \in \mathbb{C}^{Q N_{c} (L+1)-1
\times L}$ has the form\\
\[
\boldsymbol{S}_k( \tau_k)=
\left[ {\begin{array}{*{20}c}
   | & | & & | \\
   \boldsymbol{S}_{k} (\tau_k,0) &
   \boldsymbol{S}_{k}(\tau_k,1)  & \cdots &  \boldsymbol{S}_{k} (\tau_k,L-1) \\
   | & | & & | \\
\end{array}} \right]
\]
and the spreading code vector $\boldsymbol{S}_k(\tau_k,\ell)
\in \mathbb{C}^{Q N_{c}(L+1)-1 \times 1}$ is given by
\[
\boldsymbol{S}_k (\tau_k,\ell)=
\left[ {\begin{array}{c}
   \boldsymbol{0}_{Q N_{c}\ell+\tau_{k} \times 1} \\
   | \\
   \boldsymbol{s}_{k}(\tau_k,\ell) \\
   | \\
   \boldsymbol{0} \\
\end{array}} \right].
\]

 The vector $\boldsymbol{s}_k (\tau_k,\ell)$ contains the
 spreading code of user~$k$ having support
 $[\ell N_{c}T_{c}, (\ell+1) N_{c}T_{c}]$ with energy
 $\boldsymbol{s}_k^{\dag}(\tau_k,\ell)\boldsymbol{s}_k
 (\tau_k,\ell)=1$.  
  Finally, $\boldsymbol{0}_{M \times 1}$ denotes the $M \times 1$-dim. all-zero
  column vector.

The block diagonal channel matrix
$\boldsymbol{A}\in \mathbb{C}^{LK\times
  LK}$ in (1) is given by  $\boldsymbol{A} =
\mbox{diag}\{\boldsymbol{A}_1,\cdots, \boldsymbol{A}_K\}$. The channel
matrix for user~$k$,
$\boldsymbol{A}_k \in \mathbb{C}^{L\times L}$,
 is given by
$\boldsymbol{A}_k = {\bf I}_{L} \otimes$ $a_{k}$ where  $\bf{I}_{L}$ is the $L$-dim. identity matrix, and the symbol $\otimes$
denotes the Kronecker product. The $k$th user's channel coefficient
$a_{k}$ is a circularly symmetric complex
Gaussian random variable with zero mean and variance $\sigma_k^2$. The
$k$th user's transmission delay is assumed to be uniformly distributed.

The symbol vector $\boldsymbol{d} \in \mathbb{C}^{LK}$ takes the form $\boldsymbol{d} =\mbox{col}\{{\boldsymbol{d}_1,\cdots,\boldsymbol{d} _K}\}$
where the vector $\boldsymbol{d}_k  \in\mathbb{C}^{L} $ contains the
$k$th user's symbols, i.e. $\boldsymbol{d} _k =
\mbox{col}\{d_{k}(0),\cdots,d_{k}(L-1) \}$
with $d_{k}(\ell) \in
\{-1,+1\}$ denoting the symbol transmitted by the $k$th user during
the $\ell$th signalling interval. Finally, the column vector
$\boldsymbol{w} \in \mathbb{C}^{QN_{c}(L+1)-1}$ contains complex,
circularly symmetric white Gaussian noise having  covariance matrix
$N_{0}{\bf I}$.
We assume that the vectors
$\boldsymbol{a}\triangleq \mbox{col}\{a_{1},a_{2},\cdots,a_{K}\}$,
$\boldsymbol{\tau}\triangleq
\mbox{col}\{\tau_{1},\tau_{2},\cdots,\tau_{K}\}$, $\bd$ and
$\boldsymbol{w}$ and their components are independent.
The receiver does not know the data sequences, the (complex) channel coefficients, or the  transmission delays.

\section{Monte-Carlo SAGE Joint Parameter Estimation}
\label{sec:MCSAGE_appl}

\subsection{The SAGE Algorithm}

In previous applications, the SAGE algorithm \cite{fessler} has been extensively used
to iteratively approximate the ML/MAP estimate of a parameter vector  $\boldsymbol{\theta}$
with respect to the observed data $\br$.
 To obtain a receiver architecture that iterates between soft-data and
 channel estimation, one might choose the parameter vector as
 $\bTheta=\left\{\mathfrak{R}(a_{1}),\cdots,
   \mathfrak{R}(a_{K}),\mathfrak{I}(a_{1}),\cdots,\mathfrak{I}(a_{K}),\tau_{1},\cdots,\tau_{K} \right\}$. The symbols $\mathfrak{R}(\cdot)$ and $\mathfrak{I}(\cdot)$ denote the real and imaginary parts of the complex argument, respectively. At iteration $i$,
only the parameter vector of user $k$, $\bTheta_{k}$ are updated, while the
parameter vectors of the other users ${\bTheta}_{\bar k}=\bTheta
\backslash \bTheta_{k}$ are kept fixed. In the SAGE framework $\br$ is
referred to as the \emph{incomplete} data. The so-called  {\em admissible hidden} data
$\bchi_k$ with respect to $\bTheta$ is selected to be
$\bchi_k=\{\br,\bd\}$. Notice that $\bchi_k$ can only be partially
observed. Applying the SAGE algorithm to MAP parameter estimation, yields the expectation (E)-step
\begin{equation}
\label{alg:sage_estep}
Q_k(\bTheta_k,\bTheta^{[i]})=E_{\bd} \left\{ \log
  p\left(\br,\bd,\ba_k,\btau_k,{\ba}_{\bar k}^{[i]},{\btau}_{\bar k}^{[i]} \right) \mid \br,\ba^{[i]},\btau^{[i]}  \right\}.
\end{equation}
The maximization (M)-step computes a value of the argument $\btau_k$
in (\ref{alg:sage_estep}) to obtain the update
$\bTheta_k^{[i+1]}$. The objective function is non-decreasing at each
iteration.

\subsection{The Monte-Carlo SAGE algorithm}

We will see that direct computation of the expectation in
(\ref{alg:sage_estep}) requires a non-polynomial number of operations in the
number of users $K$ and thus becomes prohibitive with increasing $K$. To make the computation of the expectation in
(\ref{alg:sage_estep}) feasible though,  we propose to use the technique of Markov
chain Monte Carlo (MCMC) to obtain the Monte-Carlo SAGE
algorithm. MCMC is a statistical technique that allows generation of ergodic
pseudo-random samples $\bd^{[i,1]},\ldots,\bd^{[i,N_t]}$ from the current
approximation to the conditional pdf $p(\bd | \br, \bTheta^{[i]})$. These samples are used to approximate the expectation in
(\ref{alg:sage_estep}) by the sample-mean.  The Gibbs sampler and the
Metropolis-Hastings algorithm are widely used MCMC algorithms. Here we
describe only the Gibbs sampler \cite{Borunjeny,doucet}, as it is the
most commonly used in applications. Having initialized
$\bd^{[0,0]}$ randomly, the Gibbs sampler iterates the following loop
at SAGE iteration $i$:
\begin{itemize}
\item Draw sample $\bd_1^{[i,t]}$ from
  $p(\bd_1|\bd_2^{[i,t-1]},\ldots,\bd_K^{[i,t-1]}, \br, \bTheta^{[i]})$\\
\item Draw sample $\bd_2^{[i,t]}$ from
  $p(\bd_2|\bd_1^{[i,t]},\bd_3^{[i,t-1]}\ldots,\bd_K^{[i,t-1]}, \br,
  \bTheta^{[i]})$\\
\vdots
\item  Draw sample $\bd_K^{[i,t]}$ from
  $p(\bd_K|\bd_1^{[i,t]},\ldots,\bd_{K-1}^{[i,t]}, \br, \bTheta^{[i]})$\\
\end{itemize}
Following this approach, we have
\begin{equation*}
\label{alg:mcmc_sage_estep}
Q_k(\bTheta_k,\bTheta^{[i]})=\frac{1}{N_t} \sum_{t=1}^{N_t} \left\{ \log
  p\left(\br,\bd^{[i,t]},\ba_k,\btau_k,{\ba}_{\bar
      k}^{[i]},{\btau}_{\bar k}^{[i]} \right) \right\}.
\end{equation*}

Notice that with increasing $N_t$, the Monte-Carlo SAGE algorithm converges to the MAP
solution $\bTheta = \bTheta^\star$ up to random fluctuations around $\bTheta^\star$ \cite{Tanner-90}.

\subsection{Receiver design}

This subsection is devoted to the derivation of a receiver
architecture for joint estimation of parameters within the Monte-Carlo SAGE framework. Discarding terms
independent of $\ba$ and $\btau$, we obtain
\begin{equation}
\log p(\br,\bd,\ba,\btau) =
\log p(\br|\bd,\ba,\btau) +
\log p(\bd) +
\log p(\ba) + \log p(\btau).
\label{appl:likefkt}
\end{equation}
From (\ref{sys:received}), it follows that
\begin{equation}
\log p(\br|\ba,\btau,\bd) \varpropto
\Re{\{\br^{\dag}\boldsymbol{S}\boldsymbol{A}\bd}\} -\frac{1}{2}
\bmu(\bTheta,\bd)^{\dag}\bmu(\bTheta,\bd),
\label{appl:likelihood}
\end{equation}
where $\bmu(\bTheta,\bd) \triangleq
\sum_{k=1}^{K}\sum_{\ell=0}^{L-1}\bS_{k}(\ell,\tau_{k})a_{k}d_{k}(\ell)$
and $(.)^{\dag}$ is the conjugate transpose of the argument.
\subsubsection{The E-step}
Substituting (\ref{appl:likelihood}) into (\ref{appl:likefkt}) yields
after some algebraic manipulations for the E-step of the Monte-Carlo SAGE algorithm:
\begin{eqnarray}
\lefteqn{Q_{k}(\bTheta_{k}|\bTheta^{[i]})=}\nonumber\\
&& \hspace*{-5ex}
 \frac{2}{N_{0}} \sum_{\ell=0}^{L-1} \Re\left\{a^{*}_{k}
\Psi(\ell,\tau_{k})\right\}-\frac{L}{N_{0}}|a_{k}|^{2} -
\frac{1}{\sigma_k^2}|a_k|^2 \label{appl:Q-fct}
\end{eqnarray}
with the branch definition
\[
\Psi(\ell,\tau_{k}) \triangleq \boldsymbol{S}^{\dag}_{k}(\ell,\tau_{k})\left(\exd{k}{\ell}\br -
  \mathcal{I}_k^{[i]}(\ell)\right)
\]
and the interference term 
\begin{eqnarray*}
 \mathcal{I}_k^{[i]}(\ell) &\triangleq& \sum_{k' \neq
k}a_k'^{[i]}\bigg(\boldsymbol{S}_{k'}(\ell+1,
\tau^{[i]}_{k'})\corrd{k}{\ell}{k'}{\ell+1}\\
&&\hspace*{1ex} +
\boldsymbol{S}_{k'}(\ell,\tau^{[i]}_{k'})\corrd{k}{\ell}{k'}{\ell}\\
&&\hspace*{1ex}+\boldsymbol{S}_{k'}
(\ell-1,\tau^{[i]}_{k'})
\corrd{k}{\ell}{k'}{\ell-1}\bigg).
\end{eqnarray*}
Moreover,
\begin{equation}
\label{appl:softdata}
 \exd{k}{\ell}\triangleq \sum_{m\in
   \mathcal{S}}m P(d_{k}(\ell)=m| \br, \btau^{[i]},\ba^{[i]})
\end{equation}
and
\begin{eqnarray}
\label{appl:softcorr}
\lefteqn{\corrd{k}{\ell}{k'}{\ell'} \triangleq
\sum_{m\in \mathcal{S}}\sum_{n\in \mathcal{S}}~m~n}\nonumber \\&&
\!\!\times P(d_{k}(\ell)=m,d_{k'}(\ell')=n \mid \br,
  \btau^{[i]},\ba^{[i]}), \mbox{   for  } k'\neq k,~
\end{eqnarray}
  where $\mathcal{S}\triangleq\{-1,+1\}$ is the signal constellation
  and the lag is within range $\ell' \in \{\ell-1,\ell,\ell+1\}$.

\subsubsection{The M-step}

The M-step of the SAGE algorithm is realized by first maximizing (\ref{appl:Q-fct}) with respect to the transmission delays $\tau_k$,
\begin{equation}
\label{appl:tau_update}
\tau^{(i+1)}_{k}=\arg \max_{\tau_k} \left|\sum_{\ell=0}^{L-1}\Psi(\ell,\tau_k)\right|.
\end{equation}
Then by inserting  (\ref{appl:tau_update}) into
(\ref{appl:Q-fct}), taking derivatives with respect to the $a_k$'s,
setting the results equal to zero, and solving yields
\[
a_k^{(i+1)} = \frac{1}{L+N_0/\sigma_k^2}\sum_{\ell=0}^{L-1}\Psi(\ell,\tau^{(i+1)}_{k}).
\]

\section{Monte-Carlo Implementation to the Computation of A Posteriori Probabilities}

\subsection{Computation of the soft-data symbols in (\ref{appl:softdata})}

\label{sec:softdata}

Let $\overline{\bd_{k}(\ell)}\triangleq \bd \backslash \{d_{k}(\ell)\}$.
For notational simplicity we use $\bar{\bd}\triangleq\overline{\bd_{k}(\ell)}$ throughout this section. Then, the {\em  a
posteriori} probability of  $d_{k}(\ell)$ in (\ref{appl:softdata}) can be evaluated as

\setlength\arraycolsep{1pt}
\begin{eqnarray}
\lefteqn{P(d_{k}(\ell)=m \mid \br,
  \btau^{[i]},\ba^{[i]})}\nonumber\\&=&\sum _{\bar{\bd}}P(d_{k}(\ell)=m \mid \bar{\bd}, \br, \btau^{[i]},\ba^{[i]})~P( \bar{\bd}|\br, \btau^{[i]},\ba^{[i]}) \nonumber\\
   &\approx&\frac{1}{N_t}\sum_{t=1}^{N_t}
   P(d_{k}(\ell)=m| \bar{\bd}^{[i,t]},\br, \btau^{[i]},\ba^{[i]}).
\label{MC:APP}
\end{eqnarray}
To compute $P(d_{k}(\ell)=m| \bar{\bd}^{[i,t]},\br,
\btau^{[i]},\ba^{[i]})$ for this Markov chain Rao-Blackwellization technique, we define
 \begin{equation}
\label{MC:LLR}
 \lambda^{[i,t]}\triangleq\ln \frac{P\left(d_{k}(\ell)= +1|\bar{\bd}^{[i,t]},\br,
\btau^{[i]},\ba^{[i]}\right)}{ P\left(d_{k}(\ell)=
-1|\bar{\bd}^{[i,t]},\br, \btau^{[i]},\ba^{[i]}\right)},
 \end{equation}
For uncoded transmission, the data symbols are i.i.d. and equally
likely. Therefore, it follows from (\ref{MC:LLR}) that
\begin{equation}
\label{MC:LLR_exp}
 \lambda^{[i,t]} = \ln \frac{P( \br \mid d_{k}(\ell)= +1, \bar{\bd}^{[i,t]},
\btau^{[i]},\ba^{[i]})}{ P( \br \mid d_{k}(\ell)= -1,
\bar{\bd}^{[i,t]},\btau^{[i]},\ba^{[i]})},
\end{equation}
from which it can be easily seen that
\begin{equation*}
P\left(d_{k}(\ell)= m\mid \bar{\bd}^{(t)},\br,
\btau^{[i]},\ba^{[i]}\right)=\frac{1}{1+\exp\left(-m\lambda^{[i,t]}\right)}.
\label{MC:logapp_lambda}
\end{equation*}
From (\ref{sys:received}),  we have $p(\br|\bD)\thicksim \exp(-\frac{1}{N_{0}}|\br-\bG
\bd|^{2})$, with $\bG\triangleq\bf{S}(\btau)\bA$ and
$\bd=\mbox{col}\{\bd_{1},\bd_{2},\cdots,\bd_{K}\},$. After some
algebra (\ref{MC:LLR_exp}) can be expressed as
\begin{equation}
\label{MC:LLR_fin}
\lambda^{[i,t]}=\frac{4}{N_{0}}
\Re\left\{(\bg^{[i]}_{q})^{\dag}(\br-{\bG}^{[i]}_{\bar q} \mbox{  }
{\bd}^{[i,t]}_{\bar q})\right\},
\end{equation}
where $q\triangleq kL+\ell$, and ${\bG}_{\bar q}$ is $\bG$ with its $q$th column
$\bg_{q}$ removed. Similarly, ${\bd}_{\bar q}$ denotes the
vector $\bd$ with its $q$th component removed.

In summary, for each $k=1,2,\cdots,K$ and $\ell=0,1,\cdots,L-1$, to
estimate the {\em a posteriori} probabilities
$P(d_{k}(\ell)|\br,\btau^{[i]},\ba^{[i]})$ in (\ref{MC:APP}), the Gibbs
sampler runs over all symbols $N_{t}$ times to generate a collection
of vectors $\left\{\bar{\bd}^{[i,t]}\triangleq
\bar{\bd}^{[i,t]}_{k}(\ell)\right\}_{t=1}^{N_{t}}$ which are used
in (\ref{MC:LLR_fin}) to estimate the desired quantities.

\subsection{Computation of  the soft-value for the product of two
  data symbols in (\ref{appl:softcorr})}

Similarly, a number of random samples
$\overline{\overline{\bd}}^{{[i,t]}}\triangleq
\overline{\overline{\bd_{k,k'}(\ell')}}^{[i,t]},
t=1,2,\cdots, N_t, \ell'
\in \{-1,0,+1\}$ are drawn, using the Gibbs sampling technique,
from the joint conditional posterior distribution,
$P(\overline{\overline{\bd}} \mid \br, \btau^{[i]},\ba^{[i]}).$ Based on the samples
$\overline{\overline{\bd}}^{[i,t]}$,  $\corrd{k}{\ell}{k'}{\ell'}$ in (\ref{appl:softcorr}) can be evaluated by
\begin{eqnarray*}
\lefteqn{\corrd{k}{\ell}{k'}{\ell'}\approx
(1/N_{t})}\nonumber\\&&
\times \sum_{t=1}^{N_{t}}
\sum_{m,n\in \mathcal{S}}m n
P\left(d_{k}(\ell)=m,d_{k'}(\ell')=n \mid \overline{\overline{\bd}}^{{[i,t]}}, \br, \btau^{[i]},
\ba^{[i]}\right).
\end{eqnarray*}
We need to evaluate the probability in the expression above. Following the same route taken as in the previous section and after some algebra, it can be expressed as
\[
P\left(d_{k}(\ell)= m, d_{k'}(\ell')= n \mid \overline{\overline{\bd}}^{[i,t]}, \br, \btau^{[i]},
\ba^{[i]}\right)=\]
\begin{equation}\hspace*{3cm} \frac{1}{1+\exp\left(-\zeta^{[i,t]}\right)}\cdot
\frac{1}{1+\exp\left(-\lambda^{[i,t]}\right)}.
\label{MC:LLR_corr}
\end{equation}

The quantities $\zeta^{[i,t]}$ and $\lambda^{[i,t]}$  (\ref{MC:LLR_corr}) are given by
\begin{eqnarray*}
\zeta^{[i,t]}&=&\frac{4}{N_{0}}
\Re\left\{n(\bg^{[i]}_{p})^{\dag}(\br-{\bG}^{[i]}_{\overline{p,q}}
\mbox{  }
{\bd}^{[i,t]}_{\overline{p,q}})-mn(\bg^{[i]}_{p})^{\dag}\bg^{[i]}_{q}\right\},\\
\lambda^{[i,t]}&=&\frac{4}{N_{0}}
\Re\left\{m(\bg^{[i]}_{q})^{\dag}(\br-{\bG}^{[i]}_{\bar q} \mbox{  }
{\bd}^{[i,t]}_{\bar q})\right\},
\end{eqnarray*}
where $p\triangleq k'L+\ell'$ and $q\triangleq kL+\ell$.
${\bG}_{\overline{p,q}}$ is $\bG$ with its $p$th and $q$th columns
$\bg_{p},\bg_{q}$ removed.  Similarly, ${\bd}_{\overline{p,q}}$
denotes the vector $\bd$ with its $p$th and $q$th components
removed.

\section{Performance analysis}

\subsection{ Modified Cramer-Rao Bounds for the Estimated Parameters}
We now derive the modified Cramer-Rao lower bounds (MCRB) on the
variances of any unbiased estimates $\widehat{\bTheta}$ of the
parameter vector $\bTheta$. It is shown in \cite{VanTrees} that for $\theta_{p} \in \bTheta$, $\mbox{var}(\widehat{\theta}_{p}-\theta_{p})\geq [\bI^{-1}(\bTheta)]_{pp}$, where $\bI(\bTheta)$ is the $3K\times 3K$ Fisher information matrix whose $(p,q)$th component is defined by
\begin{equation*}
[\bI(\bTheta)]_{pq}\triangleq - E_{\br,\ba} \bigg\{\frac{\partial^{2}\ln p(\br,\ba \mid \btau)}{\partial \theta_{p}\partial \theta_{q}} \bigg\}, \mbox{   for   }p,q=1,2,\cdots,3K.
\end{equation*}
For the joint likelihood function in (\ref{appl:likelihood}),
it is shown in \cite{kay} that the Fisher information matrix can be computed by
\begin{equation}
[\bI(\bTheta)]_{pq}=
\frac{2}{N_{0}}E_{\bd}\bigg\{E_{\ba|\bd}\bigg\{\Re\bigg[
\frac{\partial \bmu^{\dag}(\bTheta,\bd)}{\partial
  \theta_{p}}\frac{\partial \bmu(\bTheta,\bd)}{\partial
  \theta_{q}}\bigg]\bigg\}\bigg\},
\label{CRB_Fisher}
\end{equation}
$p,q=1,2,\cdots, 3K.$

Taking the expectations with respect to channel coefficients $\ba$ and data $\bd$ after taking the partial derivatives in (\ref{CRB_Fisher})  with respect to $\theta_{p}$ and $\theta_{q}$, for different regions of $p$ and $q$ values,  under the assumption that the data sequences are independent and equally likely and the fact that $\bS^{\dag}(\tau_{p},\ell)\bS(\tau_{p},\ell)=1,$ for $p=1,2\cdots K; \mbox{  }\ell=0,1,\cdots,L-1$, the Fisher information matrix becomes a diagonal matrix whose $(p,p)$th component can  be evaluated as
\begin{equation}
[\bI(\bTheta)]_{pp}=\frac{2}{N_{0}}\left\{ \begin{array}{ll}
        L; & p=1,\cdots,K\\
        L; & p=K+1,\cdots,2K\\
        \sigma_{p}^{2}\sum_{\ell=0}^{L-1} \mid \bS'(\ell) \mid^{2}; & p=2K+1,\cdots,3K.\\
   \end{array}
   \right.
\label{CRB_Fisher_eval}
\end{equation}
with the short-cut $\bS'[\ell] \triangleq
\frac{\partial\bS_p(\tau_{p},\ell)}{\partial\tau_{p}} \mid_{t=\ell
  T_b + \widehat\tau_p}$.
The final result for the MCRBs on the estimates of the channel coefficients and the transmission delays is obtained by inverting the diagonal matrix $\bI(\bTheta)$ in (\ref{CRB_Fisher_eval}) as follows.
\begin{eqnarray}
\mbox{var}(\widehat{a}_{k})&\geq & N_0/L, \label{MCRB:a} \\
\mbox{  }\mbox{var}(\widehat{\tau}_{k})&\geq & 1/(8 \pi^{2} L\overline{\gamma_{k}}\mbox{  }B^{2}_{s_{k}} ) \label{MCRB:tau},
\end{eqnarray}
$k = 1,2,\dots,K$. The symbol $\overline{\gamma_{k}}\triangleq \sigma^{2}_{k}/N_{0}$
is the average SNR, $B_{s_{k}}$ is the Gabor bandwidth of the $k$th user's
spreading code waveform, $s_{k}(t)$ i.e., 
\begin{equation*}
B_{s_{k}}\triangleq \bigg(\int_{-\infty}^{+\infty}f^{2} \mid S_{k}(f)\mid^{2} df\bigg)^{1/2},
\end{equation*}
and $S_{k}(f)$ is the Fourier transform of $s_{k}(t), t\in
[0,T_{b}]$. Note that the Gabor bandwidth $B_{s_{k}}$ tends to infinity for rectangular-shaped (continuous-time)
chip waveforms.

\subsection{Numerical Examples}

To assess the performance of the proposed (non-linear) Monte-Carlo SAGE scheme, an asynchronous uncoded DS-CDMA system with $K = 5$ users, rectangular chip waveforms with processing gain $N_c = 8$, and $L = 80$ transmitted symbols per block is considered. The receiver processes $Q = 12$ samples per chip. For each data block, Gibbs sampling is performed over $50$ iterations. 
A few, say $L_p=4$ pilot symbols are embedded in each block to overcome the phase ambiguity problem. Each user's strongest path from the MMSE
estimate of $\ba$ given the $K~Q~(L_p+1)-1$ samples of $\br$ and the pilot symbols, yield the initial estimates $\ba^{[0]}$ and $\btau^{[0]}$. The MMSE estimate of $\boldsymbol{d}$, given
$\br$ and weighted by $\ba^{[0]}$ yields the initial symbol estimate
$\bd^{[0]}$. We refer to this method as MMSE-separate estimation (MMSE-SE). 
For comparison purpose, the SAGE-scheme for joint data detection and channel estimation in
\cite{kocian2007} for known transmission delays and hard-decision
decoding has also been considered subsequently. We refer to these scheme as
"SAGE-JDE, $\btau$ known".\\

\begin{figure}
\begin{center}
\begin{psfrags}
\psfrag{0.10}[][][0.8]{$0.10$}
\psfrag{0.15}[][][0.8]{$0.15$}
\psfrag{0.20}[][][0.8]{$0.20$}
\psfrag{0.25}[][][0.8]{$0.25$}
\psfrag{0.30}[][][0.8]{$0.30$}
\psfrag{0.35}[][][0.8]{$0.35$}
\psfrag{0.40}[][][0.8]{$0.40$}
\psfrag{0.45}[][][0.8]{$0.45$}
\psfrag{0.50}[][][0.8]{$0.50$}
\psfrag{delay}[][][0.8]{$\tau/T_b$}
\psfrag{10}[][][0.8]{$10$}
\psfrag{-1}[][][0.6]{$-1$}
\psfrag{-2}[][][0.6]{$-2$}
\psfrag{-3}[][][0.6]{$-3$}
\psfrag{user1}[l][l][0.8]{MCMC-SAGE, user 1}
\psfrag{user3}[l][l][0.8]{MCMC-SAGE, user 3}
\psfrag{MCRLB(1)xxxxxxxxxxxx}[l][l][0.8]{MCRB (\ref{MCRB:a}), user 1}
\psfrag{MCRLB(3)}[l][l][0.8]{MCRB (\ref{MCRB:a}), user 3}
\psfrag{MSE(a)}[][][0.8]{$\mbox{var}(\widehat{a}_k)$}
\includegraphics[width=10cm]{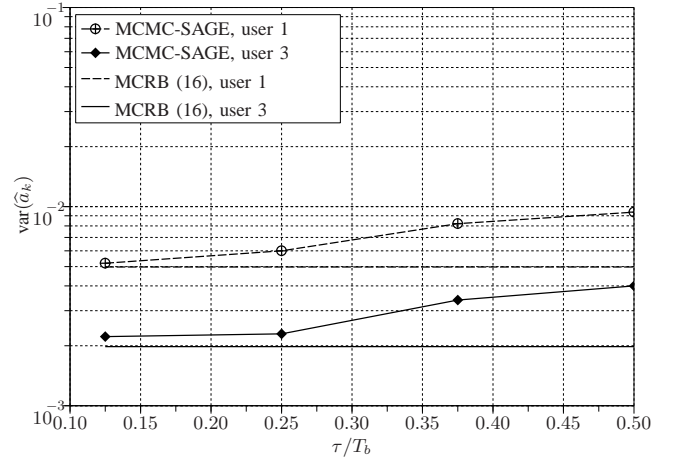}
\end{psfrags}
\end{center}
\caption{$\mbox{var}(\widehat{a}_k)$ of the MCMC-SAGE in near-far scenario. \label{fig:MMSE_a}}
\end{figure}

To study the behavior of the proposed MCMC-SAGE scheme, we consider communication over AWGN (not known to the receiver). The individual powers are given by
\begin{equation*}
\begin{array}{lll}
\sigma_1^2 = -4~\mathrm{dB}, & \hspace{2ex} \sigma_2^2 = -2~\mathrm{dB}, & \hspace{2ex} \sigma_3^2 = 0~\mathrm{dB}, \\
\sigma_4^2 = +2~\mathrm{dB}, & \hspace{2ex} \sigma_5^2 = +4~\mathrm{dB}, \\  
\end{array}
\end{equation*}
Fig.~\ref{fig:MMSE_a} shows the mean-square-error (MSE) of the channel estimates $\widehat{\boldsymbol a}_1$ (weakest user) and $\widehat{\boldsymbol a}_3$ (normal user) as a function of the normalized transmission delays $\btau/T_b$ which are uniformly distributed on the interval between zero and the value on the abscissa. It can be seen that the MCMC-SAGE performs close to the MCRB over the entire range of $\btau$. Not shown in the plot, convergence is achieved after around 25 iterations i.e., every user's parameter vector is updated five times.

\begin{figure}
\begin{center}
\begin{psfrags}
\psfrag{0.10}[][][0.8]{$0.10$}
\psfrag{0.15}[][][0.8]{$0.15$}
\psfrag{0.20}[][][0.8]{$0.20$}
\psfrag{0.25}[][][0.8]{$0.25$}
\psfrag{0.30}[][][0.8]{$0.30$}
\psfrag{0.35}[][][0.8]{$0.35$}
\psfrag{0.40}[][][0.8]{$0.40$}
\psfrag{0.45}[][][0.8]{$0.45$}
\psfrag{0.50}[][][0.8]{$0.50$}
\psfrag{delay}[][][0.8]{$\tau/T_b$}
\psfrag{10}[][][0.8]{$10$}
\psfrag{-4}[][][0.6]{$-4$}
\psfrag{-5}[][][0.6]{$-5$}
\psfrag{-6}[][][0.6]{$-6$}
\psfrag{-7}[][][0.6]{$-7$}
\psfrag{user1xxxxxxxxxxxxx}[l][l][0.8]{MCMC-SAGE, user 1}
\psfrag{user3}[l][l][0.8]{MCMC-SAGE, user 3}
\psfrag{MSE(tau)}[][][0.8]{$\mbox{var}(\widehat{\tau}_k)$}
\includegraphics[width=10cm]{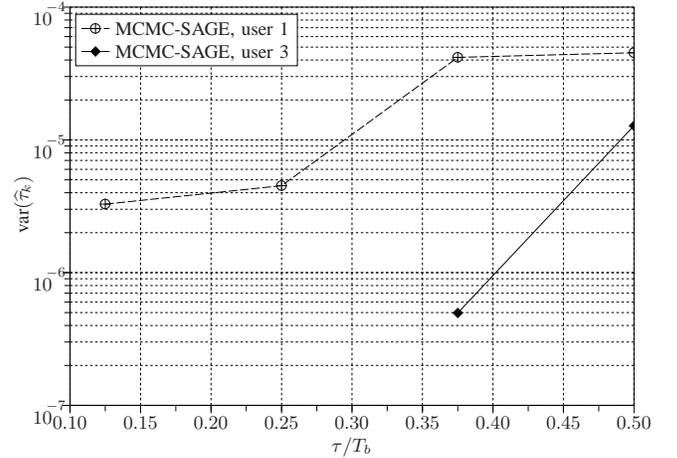}
\end{psfrags}
\end{center}
\caption{$\mbox{var}(\widehat{\tau}_k)$ of the MCMC-SAGE in near-far scenario. \label{fig:MMSE_tau}}
\end{figure}

Fig.~\ref{fig:MMSE_tau} depicts the MSE of the delay estimates $\widehat{\boldsymbol \tau}_1$ and $\widehat{\boldsymbol \tau}_3$. Notice that the MCRB for $\btau$ tends to zero for time-continuous signature waveforms. It can be seen that user~3 does not encounter delay estimation errors for small transmission delays i.e., $\tau/T_b \leq 0.2$. This effect can be partially explained by the large number of samples per chip i.e., $Q=12$. Though for higher transmission delays, $\mbox{var}(\widehat{\tau}_3)$ is finite, because of the increasing residual interference in the receiver.

\begin{figure}
\begin{center}
\begin{psfrags}
\psfrag{2}[][][0.8]{$2$}
\psfrag{4}[][][0.8]{$4$}
\psfrag{6}[][][0.8]{$6$}
\psfrag{8}[][][0.8]{$8$}
\psfrag{10}[][][0.8]{$10$}
\psfrag{12}[][][0.8]{$12$}
\psfrag{14}[][][0.8]{$14$}
\psfrag{16}[][][0.8]{$16$}
\psfrag{effective SNR}[][][0.8]{$\frac{L-L_p}{L}\bar{\gamma}_1=\frac{L-L_p}{L}\bar{\gamma}_2=\ldots=\frac{L-L_p}{L}\bar{\gamma}_K$ [dB]}
\psfrag{10}[][][0.8]{$10$}
\psfrag{0}[][][0.6]{$0$}
\psfrag{-1}[][][0.6]{$-1$}
\psfrag{-2}[][][0.6]{$-2$}
\psfrag{-3}[][][0.6]{$-3$}
\psfrag{-4}[][][0.6]{$-4$}
\psfrag{BER(user1)}[][][0.6]{${\mathrm{BER}_1}$}
\psfrag{BER(user3)}[][][0.6]{${\mathrm{BER}_3}$}
\psfrag{SU}[l][l][0.7]{SU, known channel}
\psfrag{SAGE(kc)}[l][l][0.7]{MCMC-SAGE, $\btau$ known}
\psfrag{SAGE}[l][l][0.7]{MCMC-SAGE, $\hat{\btau}$}
\psfrag{MMSE-SDEXXXXXX}[l][l][0.7]{MMSE-SE}
\includegraphics[width=10cm]{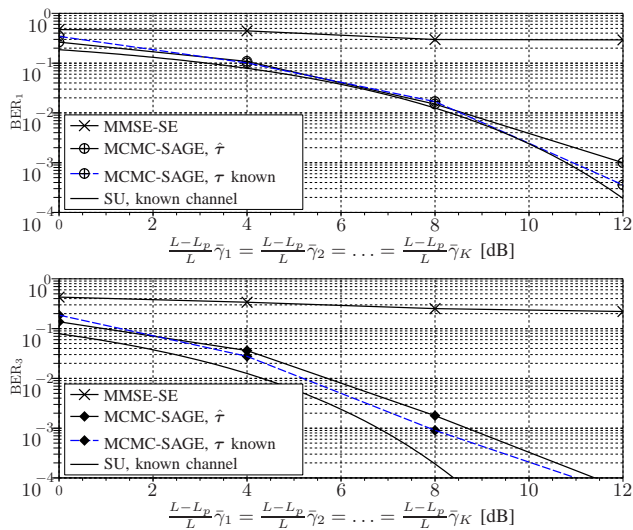}
\end{psfrags}
\end{center}
\caption{BER-performance in near-far scenario. \label{fig:BER}}
\end{figure}

The bit-error-rate ($\overline{\mathrm{BER}}$) of the proposed
receiver is plotted in Fig.~\ref{fig:BER} versus the \emph{effective}
SNR $\frac{L-L_p}{L}\bar{\gamma}_k$, $\bar{\gamma}_k \triangleq
\sigma_k^2/N_0$, $k=1,\ldots,K$. The transmission delays are uniformly distributed on $[0,T_b/2)$. It can be seen that the MMSE-SDE scheme cannot handle delay estimation errors at all due to high
correlations between the users' signature sequences. The proposed
MCMC-SAGE scheme and the "SAGE-JDE, $\btau$ known" perform similar. The weakest user 1 performs close to the single-user (SU) bound. The normal user 3 has a multiuser efficiency
of roughly 1~dB over the entire range of SNR values.

\section{Conclusions}
A computationally efficient estimation algorithm has been proposed
for  estimating the transmission delays and the channel coefficients
jointly  in a non-data-aided fashion via the SAGE algorithm. The {\em a
posteriori} probabilities needed to implement the SAGE algorithm
have been  computed by means of the Gibbs sampling technique. Exact
analytical expression have been obtained for the estimates of
transmission delays and channel coefficients. At each iteration the
likelihood function is non-decreasing.

\bibliographystyle{IEEEbib}
\bibliography{SPAWC_2009}
\end{document}